\documentclass[prd,10pt,aps,preprintnumbers,twocolumn,nofootinbib,superscriptaddress,showpacs,floatfix,amssymb,amsmath]{revtex4-1}
\usepackage{bm}
\usepackage{graphicx}
\usepackage{epstopdf}
\usepackage[cp1251]{inputenc}

\newcommand{\beqn}{\begin{eqnarray}}
\newcommand{\eeqn}{\end{eqnarray}}
\newcommand{\eq}[1]{(\ref{#1})}
\newcommand{\M}{PLSM${}_q$}
\newcommand{\cZ}{{\cal Z}}
\newcommand{\cL}{{\cal L}}
\newcommand{\dd}{{\mathrm d}}

\newcommand{\Tr}{{\mathrm{Tr}\,}}
\newcommand{\Z}{{\mathbb{Z}}}
\newcommand{\Dirac}{\rlap {\hspace{-0.5mm} \slash} D}

\begin{document}

\preprint{UUITP-01/11}

\title{Phase diagram of chirally imbalanced QCD matter}

\author{M. N. Chernodub}
\affiliation{CNRS, Laboratoire de Math\'ematiques et Physique Th\'eorique, \\ 
Universit\'e Fran\c{c}ois-Rabelais Tours, 
F\'ed\'eration Denis Poisson, Parc de Grandmont, Tours, 37200, France}
\affiliation{Department of Physics and Astronomy, University of Gent, Krijgslaan 281, S9, B-9000 Gent, Belgium}
\author{A. S. Nedelin}
\affiliation{Department of Physics and Astronomy, Uppsala University, P.O. Box 803, Uppsala, S-75108, Sweden}
\affiliation{Institute for Theoretical and Experimental Physics, B.Cheremushkinskaya 25, Moscow, Russia}

\begin{abstract}
We compute the QCD phase diagram in the plane of the chiral chemical potential and temperature using the linear sigma model coupled to quarks and to the Polyakov loop. The chiral chemical potential accounts for effects of imbalanced chirality due to QCD sphaleron transitions which may emerge in heavy-ion collisions. We found three effects caused by the chiral chemical potential: the imbalanced chirality (i) tightens the link between deconfinement and chiral phase transitions; (ii) lowers the common critical temperature; (iii) strengthens the order of the phase transition by converting the crossover into the strong first order phase transition passing via the second order end-point. Since the fermionic determinant with the chiral chemical potential has no sign problem, the chirally imbalanced QCD matter can be studied in numerical lattice simulations.
\end{abstract}


\date{February 1, 2011}

\maketitle

\section{Introduction}

It is well known that the QCD vacuum has a nontrivial topological structure due to presence of certain gluon configurations, instantons, which are characterized by an integer-valued topological winding number~\cite{ref:instanton}. For a long time an experimental evidence for the existence of the topological gluon configurations could only be found indirectly, in certain features of the meson spectrum~\cite{ref:meson}. 

Recently, it was noticed that a potentially observable direct signature of the topologically nontrivial gluon configurations can emerge in noncentral collisions of heavy ions~\cite{ref:CME:initial,ref:CME}. Such collisions create hot expanding fireballs of the quark-gluon plasma in the background of a strong magnetic field. Topologically nontrivial sphaleron transitions~\cite{ref:sphalerons,Moore:2000ara} can induce -- acting via the axial anomaly -- a nonvanishing chiral density of quarks in the plasma fireballs. This chirally imbalanced matter is characterized by different densities of right- and left-handed quarks. If such media is placed in  an external magnetic field, then an electric current of quarks should emerge along the field's axis. The unusual generation of the electric current in the magnetic field background is the essence of the so-called ``chiral magnetic effect''~\cite{ref:CME:initial,ref:CME}. Signatures of this phenomenon were found in heavy-ion experiments at the BNL Relativistic Heavy Ion Collider (RHIC) at Brookhaven~\cite{ref:RHIC} and they may also be observed in the heavy-ion collisions at the Large Hadron Collider (LHC) at CERN. 

The chiral magnetic effect is realized in the chirally imbalanced media in the strong magnetic field background. It is also important that the system is hot enough to be in the deconfining and chirally restored phase. The later requirement is essential because the system should be able to generate the quark's electric current so that the quarks should be deconfined. Moreover, in the chirally broken (low temperature) phase the chiral imbalance should quickly be washed out due to the presence of the chiral condensate which facilitates transitions between left and right quarks. Therefore, it becomes interesting to investigate the influence of the background magnetic field and the effect of the chiral imbalance on the finite-temperature transition between hadron and quark-gluon plasma phases. 

The effect of the strong magnetic field on the QCD phase transition was studied both analytically~\cite{ref:chiral,ref:mu5:NJL,ref:splitting,ref:splitting:small} and numerically~\cite{ref:lattice:magnetic}. All these studies have found that the magnetic field background increases the transition temperature and makes the phase transition stronger. A third effect was found in Ref.~\cite{ref:splitting}: the magnetic field splits the deconfining and chiral phase transitions, thus leading to emergence of the new, chirally broken deconfining phase (the splitting can be small~\cite{ref:splitting:small}, however). In the low-temperature and strong-magnetic-field corner of the QCD phase diagram a new electromagetically superconducting phase may emerge~\cite{ref:superconductor}.

The topologically induced changes in chirality can be modeled with the help of the chiral chemical potential $\mu_5$ which creates a difference between the right- and left-handed particles~\cite{ref:CME}. This potential can be related to the $\theta$ angle of strong interactions as follows~\cite{ref:CME}:
\beqn
\mu_5 = \frac{\partial}{\partial t} \frac{\theta}{2 N_f}\,,
\eeqn
where $t$ is the time coordinate and $N_f$ is the number of the light flavors in the theory.

The influence of the chirally imbalance on the thermal phase transition in the magnetic field background was addressed in Ref.~\cite{ref:mu5:NJL} regarding possible applications to the chiral magnetic effect. Working in the Nambu--Jona-Lasinio (NJL) model coupled to the Polyakov loop (PNJL), the authors of Ref.~\cite{ref:mu5:NJL} have found that the chiral imbalance makes the temperature of the chiral phase transition smaller while the strength of the transition becomes stronger. In our paper we confirm findings of Ref.~\cite{ref:mu5:NJL} working at zero magnetic field 
in the linear sigma model coupled to quarks and to the Polyakov loop (\M) which also serves as an effective low-energy model of QCD. 
Our main result is the QCD phase diagram in the $(\mu_5,T)$ plane, which is plotted in Fig.~\ref{fig:phase}. 

The structure of this paper is as follows: in Section~\ref{sec:model} we describe our model (\M), in Section~\ref{sec:phase} we discuss thermodynamics and calculate the phase diagram of the model, and Section~\ref{sec:conclusions} is devoted to our conclusions.

\section{The model}
\label{sec:model}

We use a linear sigma model coupled to quarks~\cite{ref:LSMq} and to the Polyakov loop (\M)~\cite{ref:PLSMq}. The inclusion of the Polyakov loop allows us to account for effect of the color confinement following similar proposal in the NJL model~\cite{ref:PNJL}.

This low-energy model of QCD contains three types of fields: the duplet of the quark fields $\psi(x) = (u,d)^T$, the scalar chiral (meson) fields  $(\sigma,\vec{\pi})$  with $\vec{\pi} = (\pi_{1},\pi_{2},\pi_{3})$, and the complex-valued scalar field of the Polyakov loop,
\beqn
L(x) = \frac{1}{3} \Tr \Phi(x)\,,
\quad
\Phi = {\cal P} \exp \Bigl[i \int\nolimits_0^{1/T}
\dd \tau A_4(\vec x, \tau) \Bigr]\,, \quad
\label{polyakov_loop_definition}
\eeqn
where $A_{4}=iA_{0}$ is the timelike component of the $SU(3)$ gauge field, $\mathcal{P}$ is the path-ordering operator, $\vec{\pi}$ is the isotriplet of the pseudoscalar pions, and $\sigma$ is the pseudoscalar field. 

The \M\ Lagrangian can be represented as a sum of the following three parts:
\beqn
\cL = \cL_q(\bar \psi, \psi, \sigma, {\vec \pi}, L) + \cL_\sigma(\sigma, \vec \pi) + \cL_L(L)\,.
\label{eq:LL}
\eeqn
The quark part of the Lagrangian~\eq{eq:LL}, 
\beqn
{\cL}_q =
 \overline{\psi} \left[i \Dirac - g(\sigma +i\gamma^{5} \vec{\tau} \cdot \vec{\pi}) +\mu_{5}\gamma^{0}\gamma^{5}\right]\psi\,,
\label{quark_sector_lagrangian}
\eeqn
provides the interaction between the quarks $\psi$, the chiral fields $\sigma$, $\vec \pi$, and the gauge field $A_{\mu}$ via the covariant derivative $\Dirac=\gamma^{\mu} (\partial_{\mu}-iA_{\mu})$. The Lagrangian~\eq{quark_sector_lagrangian} includes also the real-valued chiral chemical potential $\mu_{5}$. 

The dynamics of the chiral fields is described by the second term in the Lagrangian~\eq{eq:LL},
\beqn
{\cL}_{\sigma} (\sigma,\vec{\pi}) & = & \frac{1}{2}\left(\partial_{\mu}\sigma\partial^{\mu}\sigma+\partial_{\mu}\pi^{0}\partial^{\mu}\pi^{0}\right)+
\partial_{\mu}\pi^{+}\partial_{\mu}\pi^{-} \nonumber \\
& & - V_\sigma(\sigma,\vec{\pi})\,,
\label{chiral_lagrangian}
\eeqn
where we have introduced charged and neutral mesons, 
\beqn
\pi^{\pm}=\frac{1}{\sqrt{2}}\left(\pi^{1}\pm i\pi^{2}\right)\,,\qquad \pi^{0}=\pi^{3}\,,
\label{pions_definition}
\eeqn
respectively. The first and the second terms of the potential,
\beqn
V_{\sigma}(\sigma,\vec{\pi})=\frac{\lambda}{4}\left(\sigma^{2}+\vec{\pi}^{2}-{\it v}^{2}\right)^{2}-h\sigma\,,
\label{chiral_potential}
\eeqn 
provide, respectively, (strong) spontaneous and (weak) explicit breaking of the chiral symmetry. The phenomenologically acceptable parameters in Eqs.~\eq{quark_sector_lagrangian} and \eq{chiral_potential} are: $g = 3.3$, $\lambda = 20$ and $v = 87.7$\,MeV~\cite{ref:Scavenius}. We work in a mean field approximation thus neglecting quantum fluctuations (i.e., the kinetic terms) of the scalar fields $\sigma$ and $\vec{\pi}$.

Effects of the color confinement are encoded in the last term of the Lagrangian~\eq{eq:LL}, which describes the potential of the Polyakov loop:
\beqn
{\cL}_{L}=-V_{L}(L,\,T)
\label{Polyakov_loop_lagrangian}
\eeqn
As in the previous case involving the mesonic chiral fields, we neglect possible kinetic terms for the Polyakov loop and consider only the potential term following Ref.~\cite{Rossner:2007ik}:
\beqn
\label{Polyakov_loop_potentail_definition}
& & \frac{V_L(L,T)}{T^{4}} =-\frac{1}{2}a(T)L^{*}L \\
& & + b(T)\ln\left[1-6L^*L+4\left({L^{*}}^3+L^{3}\right) - 3\left(L^{*}L\right)^2\right]\,,
\nonumber
\eeqn
where
\beqn
a(T) & = & a_{0} + a_{1}\left(\frac{T_{0}}{T}\right) +a_2\left(\frac{T_{0}}{T}\right)^{2}\,,
\nonumber \\
b(T) & = &  b_{3}\left(\frac{T_{0}}{T} \right)^{3}\,,
\quad
\label{coeffucents_of_Polyakov_loop_potential}
\eeqn
and $T_0  = 270\, \mbox{MeV}$ is the temperature of the deconfinement phase transition in the pure $SU(3)$ Yang-Mills theory without quarks.
The coefficients in Eq.~(\ref{coeffucents_of_Polyakov_loop_potential}) are:
\beqn
\label{parameters_of_Polyakov_loop_potential}
\begin{array}{llllll}
a_0 & = & 16\,\pi^2/45 \approx 3.51\,, & \qquad a_1 & = & -2.47\,, \\
a_2 & = & 15.2\,,                            & \qquad b_3 & = & -1.75\,.
\end{array}
\eeqn
The Polyakov loop potential~\eq{Polyakov_loop_potentail_definition} respects the center $\Z_3$ symmetry, $L \to e^{2 \pi n i/3} L$ with $n=0,1,2$.

\section{Thermodynamics and phase diagram}
\label{sec:phase}

The free energy density of the system is:
\beqn
\Omega=-\frac{T}{V_{3d}}\ln{\cZ}\,,
\label{free_energy_defenition}
\eeqn
where ${\cZ}$ is the partition function and $V_{3d}$ is the volume of the three-dimensional space. In the mean field approximation the chiral fields and Polyakov loop field are considered as classical objects, so that the free energy can be rewritten as 
\beqn
\Omega\left(\sigma,\, \vec{\pi},\,L,\, T,\mu_5\right) & = & V_{\phi}(\sigma,\,\vec{\pi})+V_{L}\left(L,\,T\right) \nonumber\\
& & + \Omega_{q}(\sigma,\,\vec{\pi},\,L,\, T,\, \mu_5)\,,
\label{free_energy_mean_field_definition}
\eeqn
where the potential $V_{\phi}$ is given by Eq.~(\ref{chiral_potential}) and the potential $V_{L}$ is described in Eqs.~(\ref{Polyakov_loop_potentail_definition})-(\ref{parameters_of_Polyakov_loop_potential}).  In Eq.~\eq{free_energy_mean_field_definition} the quark part $\Omega_{q}$ comes from the fermion determinant, which can be computed by explicit diagonalization of the quadratic operator corresponding to the quark Lagrangian~\eq{quark_sector_lagrangian}:
\beqn
& & \Omega_{q} = - 2 \sum\limits_{s=\pm1}\int\frac{d^{3}p}{(2\pi)^{3}} \Bigl(3\omega 
\label{free_energy_through_double_logarithms}
\\
& & +  T \Bigl\{\ln \Bigl[1{+}3\left(L^*{+}Le^{- \omega_s/T}\right)
e^{-\omega_s/T}{+}e^{-3\omega_s/T}\Bigr] {+} c.c. \Bigr\} \biggr),
\nonumber
\eeqn
where $L$ is the Polyakov loop (\ref{polyakov_loop_definition}) and the fermion spectrum is
\beqn
\omega_s(p) = \sqrt{(|p| s-\mu_{5})^{2}+g^{2}( \sigma^2 +\vec \pi^{2})}\,.
\label{energy_spectrum}
\eeqn
Here $s=\pm 1$ is the helicity (the sign of the projection of the particle's spin on the direction of the particle's motion). The first term in Eq.~\eq{free_energy_through_double_logarithms} corresponds to the divergent energy of the Dirac sea. After a proper regularization the contribution from this term renormalizes the parameters $\lambda$ and $v$ of the chiral potential (\ref{chiral_potential}) adding also finite logarithmic corrections (which are sometimes called zero-point corrections) to the potential. These corrections have a pure vacuum origin because they depend neither on temperature nor on the chiral chemical potential. Usually, these logarithmic corrections are disregarded in the phenomenological approaches based on the linear sigma model because their contribution does not change the qualitative physical picture. We ignore the logarithmic corrections as well, referring the interested reader to studies of the vacuum fluctuations effects done in Refs.~\cite{ref:vacuum:logs1,ref:vacuum:logs2}.

Integrating Eq.~(\ref{free_energy_through_double_logarithms}) by parts, we can rewrite the free energy in the following form:
\beqn
\Omega_{q} = - \frac{1}{3\pi^{2}}\sum\limits_{s=\pm1}\int\nolimits^\infty_0 dp\, p^{3} \bigl[n_{q,s}(p)+n_{\bar{q},s}(p)\bigr] \frac{\partial\omega_s(p)}{\partial p}\,, \qquad
\label{free_energy_integrated_by_parts}
\eeqn
where $n_{q,s}$ and $n_{\bar q,s} \equiv n_{q,s}^*$ are, respectively, the occupation numbers (summed over colors) for quarks and antiquarks carrying the spirality $s$, and 
\beqn
n_{q,s} = \frac{3e^{- \omega_s/T} \left(L+2L^{*}e^{-\omega_s/T}+e^{-2\omega_s/T}\right)}{1+3\left(L+L^{*}e^{-\omega_s/T}\right)e^{-\omega_s/T}+ e^{-3\omega_s/T}}\,, \qquad
\label{quark_occupation_number}
\eeqn

The mean field values of the fields $\sigma$, $\vec{\pi}$ and $L$ are found by a (numerical) minimization of the free energy given by Eqs.~(\ref{free_energy_mean_field_definition}), (\ref{chiral_potential}), (\ref{Polyakov_loop_potentail_definition})-(\ref{parameters_of_Polyakov_loop_potential}), and (\ref{free_energy_integrated_by_parts}) with respect to the variations of these fields at fixed values of the temperature $T$ and chiral chemical potential $\mu_5$. 

The fermion determinant breaks the center $\Z_3$ symmetry of the Polyakov loop potential and the global minimum of the free energy~\eq{free_energy_mean_field_definition} corresponds to the real-valued Polyakov loop. Moreover, one can show numerically that the presence of the charged pion condensates $\pi^\pm$ makes the free energy larger so that these condensates are disfavored, $\langle \pi^\pm \rangle = 0$. The last term in Eq.~\eq{chiral_potential} forces the neutral pion condensate to be zero, $\langle \pi^0 \rangle = 0$. 

Therefore we are left with two unknown expectation values  which are to be fixed by the minimization of the free energy. These are the real part of the Polyakov loop, ${\mathrm{Re}}\,L$, and the chiral order parameter $\sigma$.

\begin{figure}[!htb]
\begin{center}
\includegraphics[width=77.8mm,clip=true]{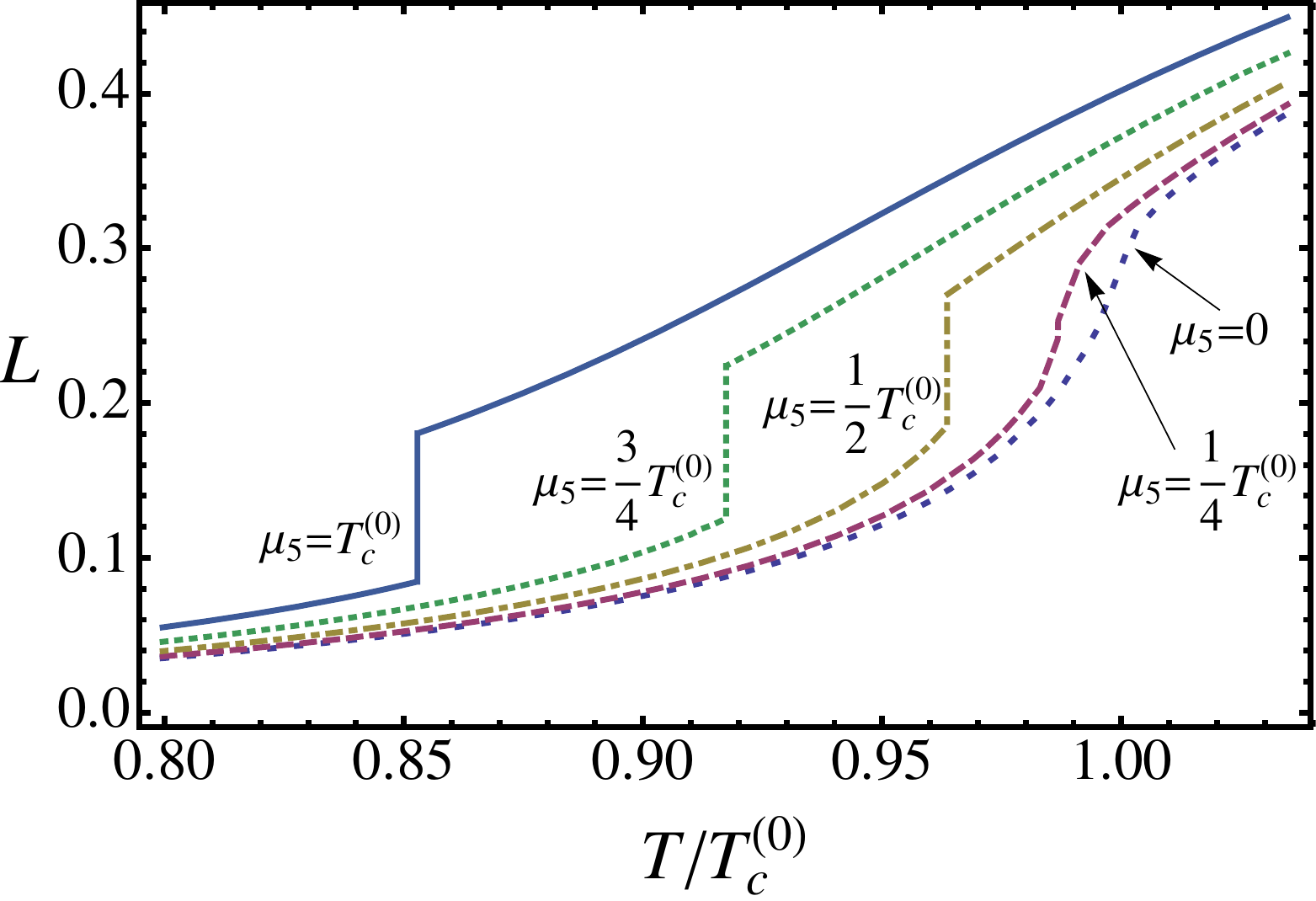} 
\end{center}
\caption{The expectation value of the Polyakov loop as the function of temperature $T$ at fixed values of the chiral chemical potential $\mu_5$ [$T$ and $\mu_5$ are given in units of $T_c^{(0)}$, the critical temperature of the $\mu_5 = 0$ transition, Eq.~\eq{eq:Tc0}].}
\label{fig:L}
\end{figure}

\begin{figure}[!htb]
\begin{center}
\includegraphics[width=79.5mm,clip=true]{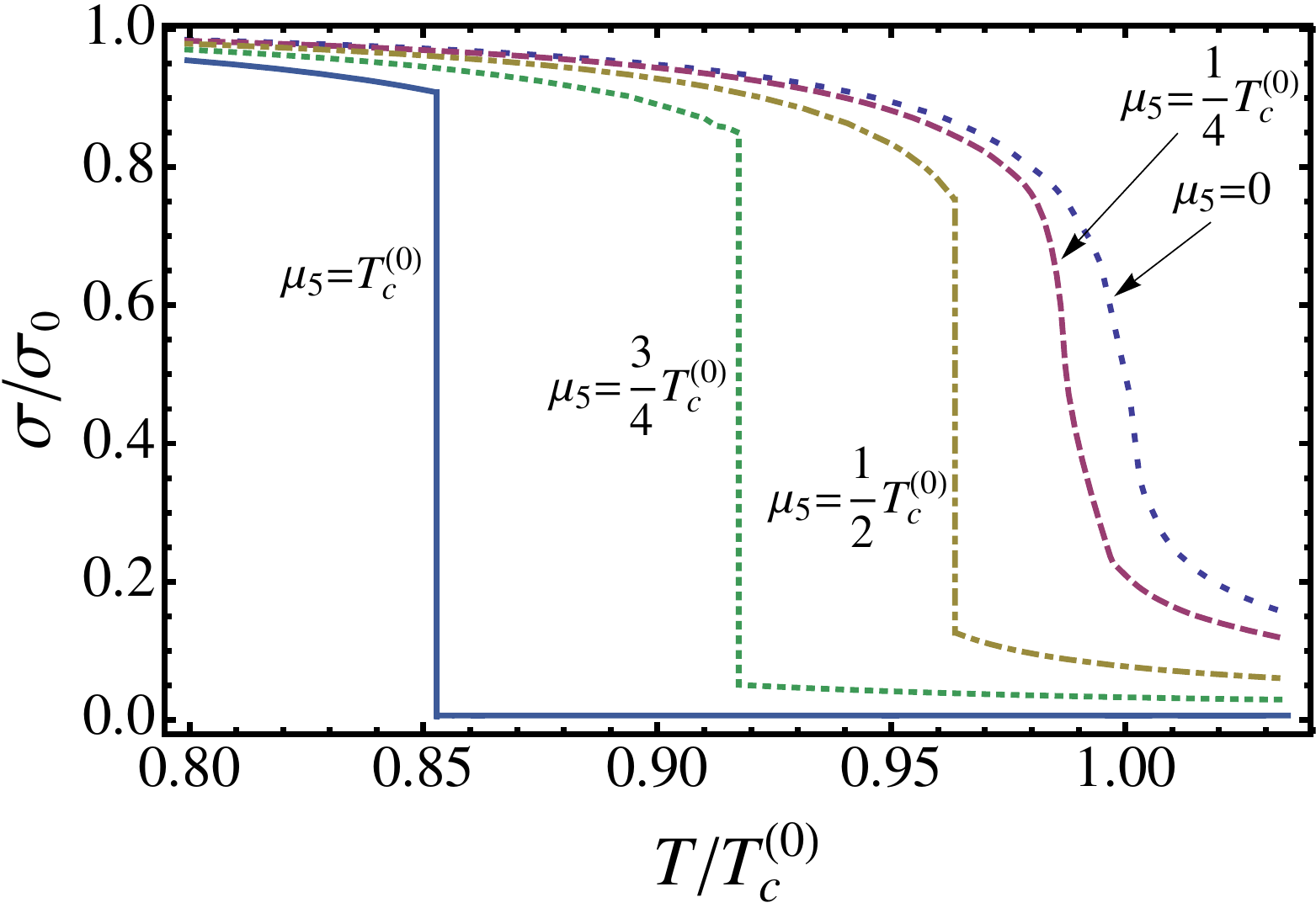}
\end{center}
\caption{The chiral order parameter $\sigma$ (we use the same notations as in Fig.~\ref{fig:L}). The value of $\sigma_0$ is given in Eq.~\eq{eq:sigma0}.}
\label{fig:sigma}
\end{figure}

The expectation values of the Polyakov loop and the chiral order parameter $\sigma$ are shown as functions of the temperature in Fig.~\ref{fig:L} and Fig.~\ref{fig:sigma}, respectively. The chiral field is normalized to unity at $T=\mu_5=0$, and the corresponding vacuum expectation value $\sigma_0$ is given by the pion decay constant:
\beqn
\sigma_0 \equiv \langle \sigma \rangle_{T=\mu_5=0} \equiv f_\pi(T{=}\mu_5{=}0) = 92.2\ \mbox{MeV}\,. \quad
\label{eq:sigma0}
\eeqn
The temperature and chiral chemical potential in Figs.~\ref{fig:L} and \ref{fig:sigma} are expressed in units of the critical transition temperature $T_c^{(0)}$ at zero chiral chemical potential in \M:
\beqn
T_c^{(0)} \equiv T_c(\mu_5=0) = 213.2\ \mbox{MeV} \qquad \mbox{[in \M]}\,. \quad
\label{eq:Tc0}
\eeqn

Due to the presence of the explicit symmetry breaking term in the chiral potential $V_\sigma$, Eq.~\eq{chiral_potential}, the transition between the quark-gluon plasma phase and the hadron phase at zero chemical potential is a smooth crossover. The critical temperature value in Eq.~\eq{eq:Tc0} corresponds to the temperature where the slopes of the Polyakov loop $L$ and the chiral field $\sigma$ are steepest\footnote{The difference in the critical values for these two quantities is smaller then one MeV in \M.}. 

\vskip 5mm

According to Figs.~\ref{fig:L} and \ref{fig:sigma}, the crossover turns into the first order transition as the chiral chemical potential $\mu_5$ increases (the larger $\mu_5$, the stronger the transition). The critical temperature of the transition between the quark-gluon plasma phase and the hadron phase decreases as the chiral chemical potential increases. Moreover, the chiral and deconfinement phase transitions are tightened to each other and they do not split as the value of $\mu_5$ increases.  

Obviously, the presence of the nonzero chiral chemical potential $\mu_5$ does not induce a nonvanishing baryon charge, so that the quark density $n$ is always zero,
\beqn
n = 2 \sum\limits_{s=\pm1} \int\frac{d^{3}p}{(2\pi)^{3}}\left[n_{q,s}(p) - n_{\bar{q,s}(p) }\right] \equiv 0\,.
\label{net_quark_density}
\eeqn 
However, the density of the chiral charge (i.e., the difference between the densities of the right- and left-handed particles),
\beqn
n_5 = n_R - n_L = \frac{\left\langle N_{5}\right\rangle}{V_{3d}}= \left\langle {\bar\psi} \gamma^{0}\gamma^{5} \psi \right\rangle = - \frac{\partial \Omega_q}{\partial \mu_5}\,,
\quad
\label{chiral_quark_density}
\eeqn 
should in general be nonzero at a nonvanishing chiral chemical potential (and, naturally, $n_5=0$ at $\mu_5=0$).

\begin{figure}[thb]
\begin{center}
\includegraphics[width=82mm,clip=true]{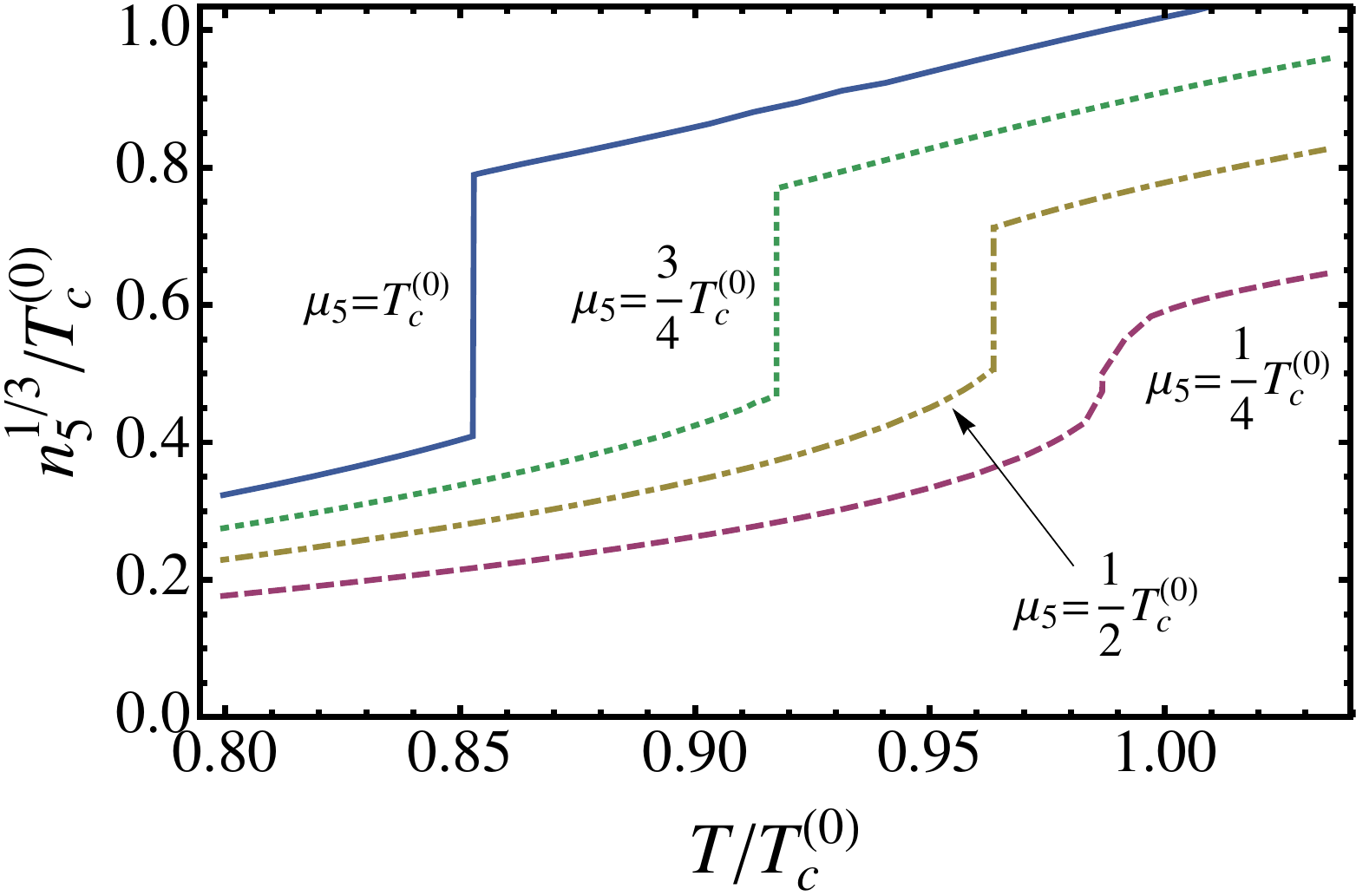}
\end{center}
\caption{The one-third power of the chiral charge density $n_5$, Eq.~\eq{chiral_quark_density}, as a function of temperature $T$ at a fixed set of the chiral chemical potential $\mu_5$ (we use the same units as in Fig.~\ref{fig:L}).}
\label{fig:n5}
\end{figure}

We plot (the one-third power of) the density of the chiral charge~\eq{chiral_quark_density} is Fig.~\ref{fig:n5}.  The chiral imbalance of the media in the presence of a fixed chemical potential $\mu_5$ is a growing function of the temperature. The chiral charge density $n_5$ is strongly enhanced at the transition temperature $T_c = T_c(\mu_5)$, as the system goes from the hadron phase to the quark-gluon plasma phase.

\begin{figure}[!htb]
\vskip 3mm
\includegraphics[width=83mm,clip=true]{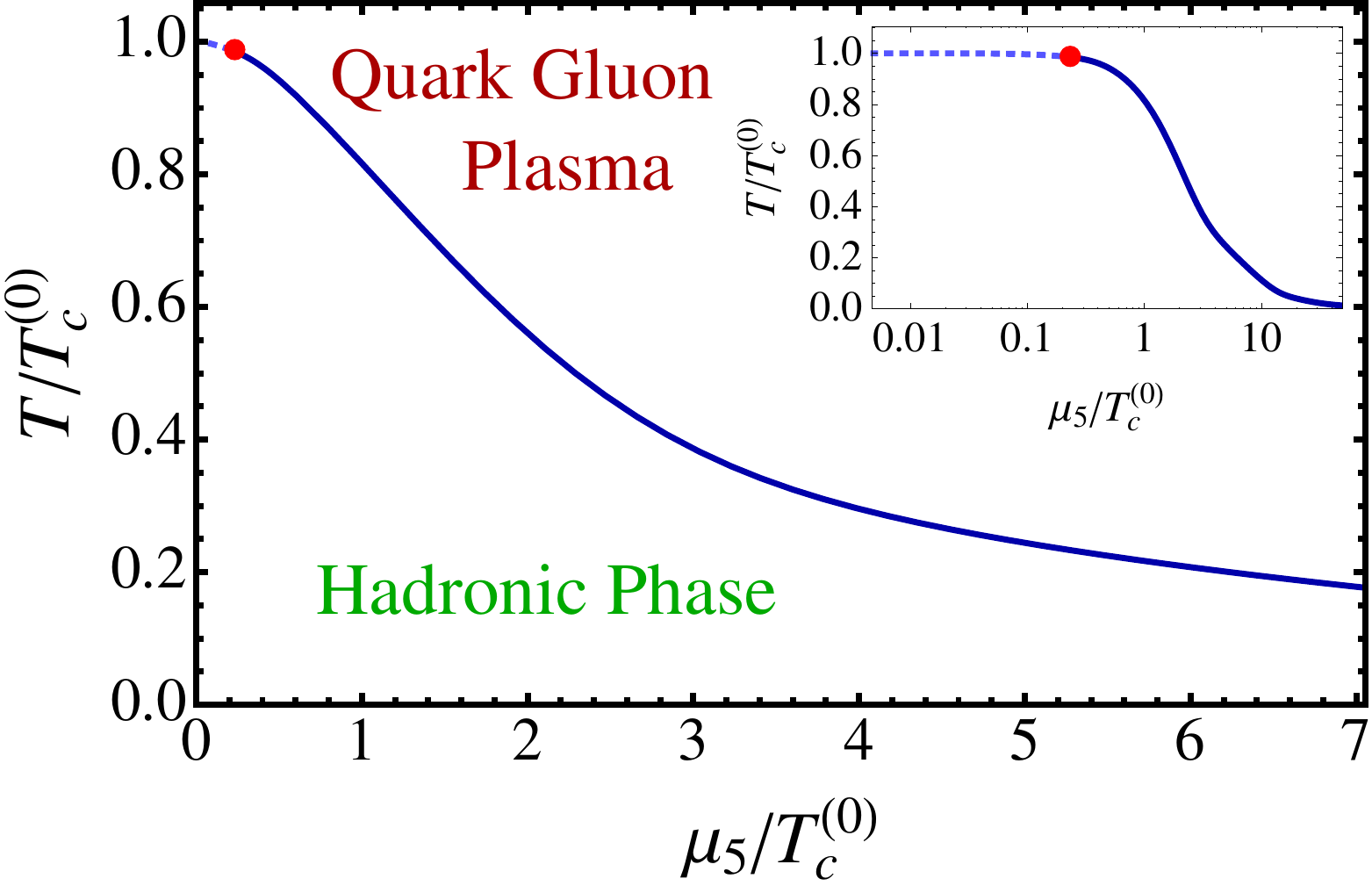}
\caption{Phase diagram of QCD in the plane of the chiral chemical potential $\mu_5$ and temperature $T$ [expressed in units of the critical transition temperature $T_c^{(0)}$ at $\mu_5=0$, Eq.~\eq{eq:Tc0}]. The dashed line represents the smooth crossover, the solid line corresponds to the first order phase transition, and the red point marks the second-order critical endpoint~\eq{eq:CEP}. The inset shows the same phase diagram with the chiral chemical potential axis plotted in the logarithmic scale.}
\label{fig:phase}
\end{figure}

The phase diagram in the $(\mu_5,T)$ plane is shown in Fig.~\ref{fig:phase}. 
The transformation of the crossover transition (the dashed line at lower values of $\mu_5$) to the first order phase transition (the solid line at larger $\mu_5$) goes via a critical endpoint (CEP) in which the transition becomes a second order transition. We found that the CEP (the red point in Fig.~\ref{fig:phase}) is located at
\beqn
(\mu_5,T)^{\mathrm{CEP}} = (0.232\, T_c^{(0)}, 0.998\, T_c^{(0)})\,.
\label{eq:CEP}
\eeqn

Our phase diagram, Fig.~\ref{fig:phase}, computed in the \M\ at zero magnetic field agrees qualitatively with the corresponding phase diagram of Ref.~\cite{ref:mu5:NJL}, where the calculations are done in the PNJL model in a weak magnetic field background\footnote{The critical temperature curve $T_c = T_c(\mu_5,B)$ is almost independent on the strength of the magnetic field $B$ in a weak field~\cite{ref:mu5:NJL}.} in the window $0 \leq \mu_5 \lesssim 2.3 \,T_c^{(0)}$. The critical curves $T_c = T_c(\mu_5)$ in the \M\ and  PNJL models are quite close to each other, while the positions of the corresponding CEP's differ from each other substantially, as the location of the CEP in the PNJL model is $(\mu_5,T)^{\mathrm{CEP}} = (1.7\, T_c^{(0)}, 0.8\, T_c^{(0)})$ according to Ref.~\cite{ref:mu5:NJL}. The difference between the PNJL prediction and our result~\eq{eq:CEP} can be attributed either to the model-dependent issues or to the fact that we have neglected the vacuum corrections coming from the fermionic determinant which may influence the \M\ thermodynamics quantitatively~\cite{ref:vacuum:logs2}.

It is important to mention that the line of the first order phase transition does not hit the $\mu_5$ axis at any finite value of the chiral chemical potential. Instead, the \M\ model predicts that the $T_c = T_c(\mu_5)$ curve approaches this axis smoothly at asymptotically large values of the chiral chemical potential $\mu_5$:
\beqn
\lim\limits_{\mu_5 \to \infty} T_c(\mu_5) = 0\,.
\label{eq:mu5inf}
\eeqn
Notice, however, that \M\ is an effective low-energy model of QCD, and therefore at large values of massive parameters (for example, at large $\mu_5$) the results coming from this model may become inaccurate. Therefore the prediction in Eq.~\eq{eq:mu5inf} should be considered with care.

\section{Conclusions}
\label{sec:conclusions}

The chirally imbalanced hot quark-gluon plasma may emerge in heavy-ion collisions at RHIC and LHC experimental facilities. We have computed the QCD phase diagram in the chirally imbalanced background at finite temperature (Fig.~\ref{fig:phase}) using the linear sigma model coupled to quarks and to the Polyakov loop (\M) at zero magnetic field. Our results are in a qualitative agreement with conclusions of an earlier study of the QCD phase finite-temperature phase transition in the PNJL model in simultaneously imposed chiral and magnetic-field backgrounds~\cite{ref:mu5:NJL}. 

We have found that the increase of the chiral chemical potential $\mu_5$ tightens the link between deconfinement and chiral phase transitions, simultaneously lowering the common critical temperature and strengthening the order of the phase transition by converting the crossover (realized at low chiral imbalance) into the first order phase transition (found at higher chiral imbalance). The location of the second-order critical endpoint -- at which the crossover turns into the first order phase transition -- is given in Eq.~\eq{eq:CEP}. The \M\ predicts that at strictly zero temperature the system stays always in the hadronic phase regardless of the value of the chiral chemical potential [Eq.~\eq{eq:mu5inf} and Fig.~\ref{fig:phase}].

We are mainly interested in the finite-temperature phase diagram of QCD at nonzero chiral chemical potential because the chiral magnetic effect is realized in the presence of the strong magnetic field in the chirally imbalanced background~\cite{ref:CME}. We confirm that the some effects of the strong magnetic field and the chiral chemical potential on the critical temperature of the QCD phase transition are opposite: the magnetic field tends to increase the temperature of the phase transition~\cite{ref:chiral,ref:mu5:NJL,ref:splitting,ref:splitting:small,ref:lattice:magnetic} while the chirally imbalanced background forces the critical temperature to become lower~\cite{ref:mu5:NJL}. Moreover, the magnetic field background splits the chiral and deconfinement phase transitions~\cite{ref:splitting} (the splitting can be small, however~\cite{ref:splitting:small}), while the chiral imbalance tightens the link between these transitions. However, despite these dissimilarities, the external magnetic field and the chiral imbalance have one common feature: they both make the QCD phase transition stronger. 

Finally, we would like to notice that the predictions for the QCD phase diagram in the $(\mu_5,T)$ plane, given in Fig.~\ref{fig:phase} of the present article and in Ref.~\cite{ref:mu5:NJL}, can directly be tested from the first principles in numerical simulations of lattice QCD because the fermionic determinant with the chiral chemical potential has no sign problem~\cite{ref:CME}.

\acknowledgments

The authors are very grateful to M.~Ruggieri for interesting discussions and useful comments. This work was partially supported by the grant ANR-10-JCJC-0408 HYPERMAG (France) and by a STINT Institutional grant IG2004-2 025 (Sweden).


\begin{thebibliography}{99}

\bibitem{ref:instanton}
  A.~A.~Belavin, A.~M.~Polyakov, A.~S.~Schwartz and Yu.~S.~Tyupkin,
  Phys.\ Lett.\  B {\bf 59}, 85 (1975).

\bibitem{ref:meson}
  E.~Witten,
  Nucl.\ Phys.\  B {\bf 156}, 269 (1979);
  G.~Veneziano,
  Nucl.\ Phys.\  B {\bf 159}, 213 (1979).
  
\bibitem{ref:CME:initial}
  D.~Kharzeev,
  Phys.\ Lett.\  B {\bf 633}, 260 (2006)
  [arXiv:hep-ph/0406125].
  
\bibitem{ref:CME}
  K.~Fukushima, D.~E.~Kharzeev and H.~J.~Warringa,
  Phys.\ Rev.\  D {\bf 78}, 074033 (2008).

\bibitem{ref:sphalerons}
  N.~S.~Manton,
  Phys.\ Rev.\  D {\bf 28}, 2019 (1983);
  F.~R.~Klinkhamer and N.~S.~Manton,
  Phys.\ Rev.\  D {\bf 30}, 2212 (1984).
 
\bibitem{Moore:2000ara}
  G.~D.~Moore,
  arXiv:hep-ph/0009161.
 
\bibitem{ref:RHIC}
  B.~I.~Abelev {\it et al.}  [STAR Collaboration],
  Phys.\ Rev.\ Lett.\  {\bf 103}, 251601 (2009)
  [arXiv:0909.1739 [nucl-ex]];
  Phys.\ Rev.\  C {\bf 81}, 054908 (2010)
  [arXiv:0909.1717 [nucl-ex]].

\bibitem{ref:chiral}
  E.~S.~Fraga and A.~J.~Mizher,
  Phys.\ Rev.\  D {\bf 78}, 025016 (2008)
  [arXiv:0804.1452 [hep-ph]].

\bibitem{ref:mu5:NJL}
  K.~Fukushima, M.~Ruggieri and R.~Gatto,
  Phys.\ Rev.\  D {\bf 81}, 114031 (2010) [arXiv:1003.0047 [hep-ph]].
  
\bibitem{ref:splitting}
  A.~J.~Mizher, M.~N.~Chernodub and E.~S.~Fraga,
  Phys.\ Rev.\  D {\bf 82}, 105016 (2010) [arXiv:1004.2712 [hep-ph]];
  R.~Gatto and M.~Ruggieri,
  Phys.\ Rev.\  D {\bf 82}, 054027 (2010) [arXiv:1007.0790 [hep-ph]].

\bibitem{ref:splitting:small}
  R.~Gatto and M.~Ruggieri,
  arXiv:1012.1291 [hep-ph].

\bibitem{ref:lattice:magnetic}
  M.~D'Elia, S.~Mukherjee and F.~Sanfilippo,
  Phys.\ Rev.\  D {\bf 82}, 051501 (2010)
  [arXiv:1005.5365 [hep-lat]].
  
\bibitem{ref:superconductor}
M.~N.~Chernodub,
  Phys.\ Rev.\  D {\bf 82}, 085011 (2010)
  [arXiv:1008.1055 [hep-ph]];
  arXiv:1101.0117 [hep-ph].

\bibitem{ref:LSMq} 
  M.~Gell-Mann and M.~Levy,
  Nuovo Cim.\  {\bf 16}, 705 (1960).
  
\bibitem{ref:PLSMq} 
  E.~Megias, E.~Ruiz Arriola and L.~L.~Salcedo,
  Phys.\ Rev.\  D {\bf 74}, 065005 (2006)
  [arXiv:hep-ph/0412308].
  
\bibitem{ref:PNJL}
  K.~Fukushima,
  Phys.\ Lett.\  B {\bf 591}, 277 (2004)
  [arXiv:hep-ph/0310121];
  C.~Ratti, M.~A.~Thaler and W.~Weise,
  Phys.\ Rev.\  D {\bf 73}, 014019 (2006)
  [arXiv:hep-ph/0506234].

\bibitem{ref:Scavenius}
  O.~Scavenius, A.~Mocsy, I.~N.~Mishustin and D.~H.~Rischke,
  Phys.\ Rev.\  C {\bf 64}, 045202 (2001)
  [arXiv:nucl-th/0007030].

\bibitem{Rossner:2007ik}
  S.~Roessner, T.~Hell, C.~Ratti and W.~Weise,
  Nucl.\ Phys.\  A {\bf 814}, 118 (2008)
  [arXiv:0712.3152 [hep-ph]].

\bibitem{ref:vacuum:logs1}
 A.~Mocsy, I.~N.~Mishustin, and P.~J.~Ellis,
  Phys.\ Rev.\  {\bf C70}, 015204 (2004) [nucl-th/0402070];
  J.~K.~Boomsma, and D.~Boer,
  Phys.\ Rev.\  {\bf D80}, 034019 (2009) [arXiv:0905.4660 [hep-ph]];
   T.~K.~Herbst, J.~M.~Pawlowski and B.~J.~Schaefer,
  Phys.\ Lett.\  B {\bf 696}, 58 (2011)
  [arXiv:1008.0081 [hep-ph]].

\bibitem{ref:vacuum:logs2}
  V.~Skokov, B.~Stokic, B.~Friman and K.~Redlich,
  Phys.\ Rev.\  C {\bf 82}, 015206 (2010) [arXiv:1004.2665 [hep-ph]].

\end{thebibliography}
\end{document}